\documentclass[reprint,superscriptaddress,amsmath,amssymb,aps,prl,floatfix]
{revtex4-1}
\usepackage{graphicx}
\usepackage{bm}
\usepackage{hyperref}

\pdfpageattr{/Group <</S /Transparency /I true /CS /DeviceRGB>>}

\pdfoutput=1

\begin{document}

\title{Resistivity in the Vicinity of a Van Hove Singularity: 
\texorpdfstring{Sr$_2$RuO$_4$}{Sr2RuO4} Under Uniaxial Pressure}

\author{M.\,E. Barber}
\email{barber@cpfs.mpg.de}
\affiliation{Scottish Universities Physics Alliance, School of Physics and 
Astronomy, University of St.\ Andrews, St.\ Andrews KY16 9SS, U.K.}
\affiliation{Max Planck Institute for Chemical Physics of Solids, 
N{\"o}thnitzer Str.\ 40, 01187 Dresden, Germany}
\author{A.\,S. Gibbs}
\altaffiliation[Present address: ]{ISIS Facility, Rutherford Appleton 
Laboratory, Chilton, Didcot OX11 OQX, U.K.}
\affiliation{Scottish Universities Physics Alliance, School of Physics and 
Astronomy, University of St.\ Andrews, St.\ Andrews KY16 9SS, U.K.}
\author{Y. Maeno}
\affiliation{Department of Physics, Graduate School of Science, Kyoto 
University, Kyoto 606-8502, Japan}
\author{A.\,P. Mackenzie}
\email{andy.mackenzie@cpfs.mpg.de}
\affiliation{Scottish Universities Physics Alliance, School of Physics and 
Astronomy, University of St.\ Andrews, St.\ Andrews KY16 9SS, U.K.}
\affiliation{Max Planck Institute for Chemical Physics of Solids, 
N{\"o}thnitzer Str.\ 40, 01187 Dresden, Germany}
\author{C.\,W. Hicks}
\email{hicks@cpfs.mpg.de}
\affiliation{Max Planck Institute for Chemical Physics of Solids, 
N{\"o}thnitzer Str.\ 40, 01187 Dresden, Germany}

\date{\today}

\begin{abstract}
We report the results of a combined study of the normal state resistivity and 
superconducting transition temperature $T_c$ of the unconventional 
superconductor Sr$_2$RuO$_4$ under uniaxial pressure.
There is strong evidence that as well as driving $T_c$ through a maximum at 
$\sim$3.5~K, compressive strains $\varepsilon$ of nearly 1~\% along the 
crystallographic [100] axis drive the $\gamma$ Fermi surface sheet through a 
Van Hove singularity, changing the temperature dependence of the resistivity 
from $T^2$ above and below the transition region to $T^{1.5}$ within it.
This occurs in extremely pure single crystals in which the impurity 
contribution to the resistivity is $<$100~n$\Omega$\,cm, so our study also 
highlights the potential of uniaxial pressure as a more general probe of this 
class of physics in clean systems.
\end{abstract}

\maketitle

When the shape or filling of a Fermi surface is changed such that it changes 
either the way that it connects in momentum ($k$) space or disappears 
altogether, its host metal is said to have undergone a Lifshitz 
transition~\cite{Lifshitz1960}.
This zero temperature transition has no associated local Landau order 
parameter, and is in fact one of the first identified examples in condensed 
matter physics of a topological transition.
Lifshitz transitions usually involve traversing Van Hove singularities (VHS).
These are points or, in the presence of interactions, regions of $k$-space 
associated in two-dimensional systems with divergences in the density of 
states~\cite{Volovik2017}.
Lifshitz transitions are therefore often associated with formation or 
strengthening of ordered states, with 
superconductivity~\cite{Tsuei1990,Markiewicz1997,Liu2010,Quan2016} and 
magnetism~\cite{Sarrao2007,Rodriguez2013} among the most prominently studied 
examples. 
They are also expected, even in the absence of order, to affect the electrical 
transport~\cite{Markiewicz1997, Hlubina1996, Varlamov1989}.

Although of considerable interest theoretically, experimentally tuning 
materials to Lifshitz transitions is challenging.
In zero applied magnetic field it usually requires non-stoichiometric doping to 
change the band filling.
In practice, this introduces disorder, which always complicates the 
understanding of the observed behaviour.
Magnetic field enables clean tuning by changing band energies for opposite 
spins through the Zeeman term, and has been used to good effect in the study of 
a number of 
systems~\cite{Yelland2011,Pfau2013,Aoki2016,Bastien2016,Grachtrup2017,Pfau2016}.
However, unless the intrinsic bandwidths are already very small, large fields 
are required to reach Lifshitz transitions, and magnetic fields also couple 
either constructively or destructively to many forms of order.

\begin{figure}[t]
  \centering
  \includegraphics{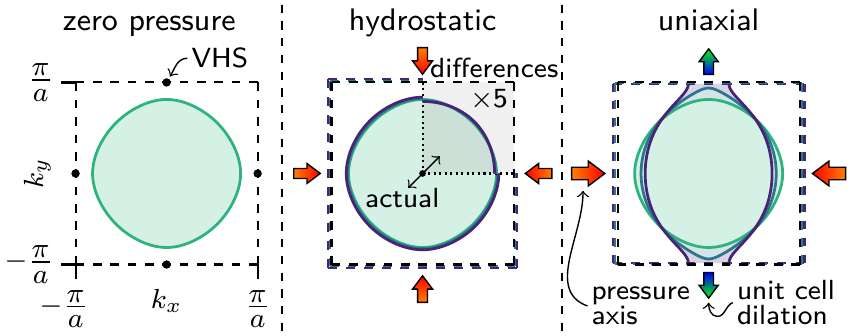}
  \caption{\label{Fig1}(color online). An illustrative single band 
tight-binding model depicting the changes of a two-dimensional Fermi surface 
under different forms of pressure.
  In general, hydrostatic pressure increases the relative weight of the 
next-nearest neighbour hopping term causing the Fermi surface to become 
slightly more circular~\cite{Burganov2016}.
  Under equal uniaxial pressures much larger distortions occur changing the 
Fermi surface from a closed to an open orbit by traversing the VHS.
  Simulation parameters are given in~\cite{Supplementary}.
  }
\end{figure}

Although high hydrostatic pressure is an option~\cite{Chu1970}, it needs to be 
strong enough either to change the relative band filling in multi-band 
materials or to substantially change the shape of a Fermi surface.
As we illustrate in Fig.~\ref{Fig1}, uniaxial pressure is in principle better 
suited to changing the shape of Fermi surfaces without the need to change the 
carrier number.
For a single band Fermi surface in two dimensions, the shape change introduced 
by applying hydrostatic pressure is negligible, while similar levels of 
uniaxial pressure introduce a large distortion.
Uniaxial pressure is therefore particularly well suited to tuning to the class 
of Lifshitz transition involving a topological change from a closed to an open 
Fermi surface by traversing the VHS associated with saddle points along one 
direction of $k$ space.
It was used a long time ago in experiments tensioning single crystal whiskers 
of the three-dimensional superconductors aluminium~\cite{Overcash1981} and 
cadmium~\cite{Watlington1977} but traversing the transition had only a weak 
effect; $T_c$, for example, changed by only $\sim$20~mK.

\begin{figure}[t]
  \centering
  \includegraphics{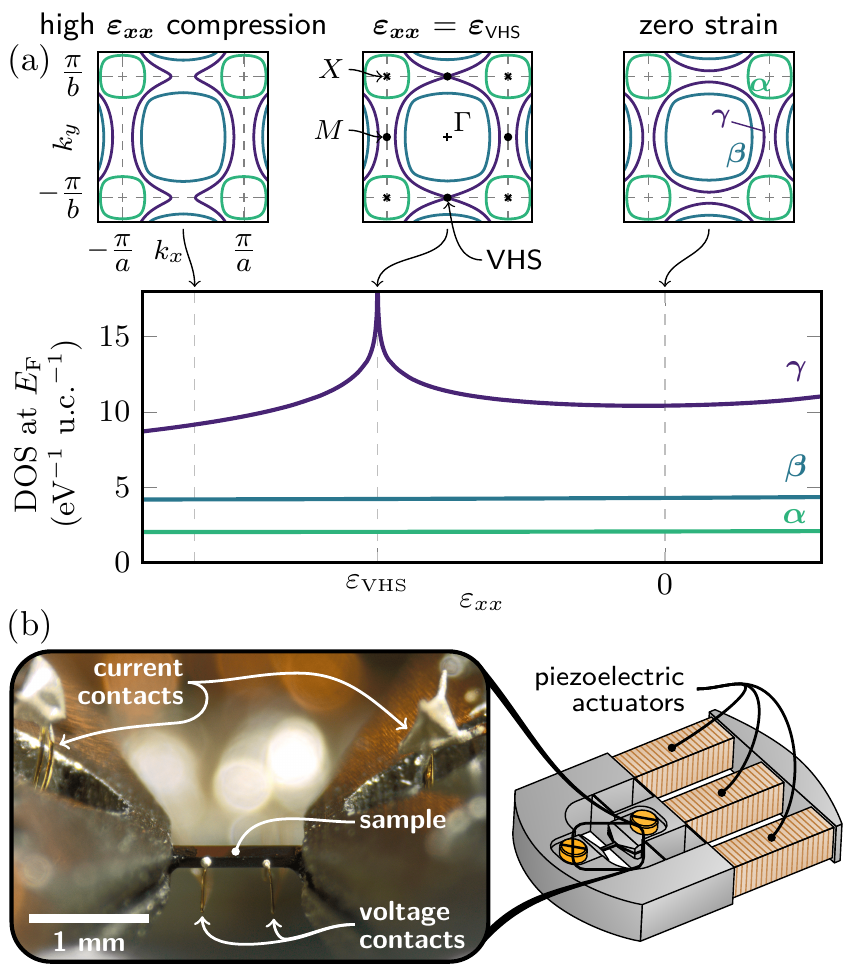}
  \caption{\label{Fig2}(color online). (a) Sr$_2$RuO$_4$ Fermi surface and 
density of states at the Fermi level as a function of applied anisotropic 
strain, calculated using a tight-binding model derived from the experimentally 
determined Fermi surface at ambient pressure~\cite{Bergemann2003} after 
introducing the simplest strain dependence for the hopping terms.
  See~\cite{Supplementary} for further simulation details.
  Fermi surfaces at three representative compressions highlight the Lifshitz 
transition as the $\gamma$-band reaches the VHS.
  (b) A sample mounted for resistivity measurements under uniaxial pressure and 
a schematic of the piezoelectric-based device used for generating the pressure.
  }
\end{figure}

Recently, we have developed novel methods of applying high levels of uniaxial 
pressure to single crystals that are not restricted to whiskers and are well 
suited to studying the more interesting case of materials with quasi-2D 
electronic structure~\cite{Hicks2014,Hicks2014_2}.
In a multi-band metal, it is possible for one of the Fermi surface sheets of a 
pressured crystal to undergo a large shape change while others are affected 
much less strongly.
As shown by the modelled Fermi surfaces in Fig.~\ref{Fig2}(a), this is the case 
for the quasi-2D material studied in this paper, 
Sr$_2$RuO$_4$~\cite{Maeno1994,Mackenzie2003,Maeno2012,Kallin2012,Mackenzie2017},
 which is predicted by first-principles calculations to undergo a Lifshitz 
transition when the lattice is compressed by $\sim$0.75~\% along a $\langle 100 
\rangle$ lattice direction~\cite{Steppke2017}.
In previous work on this material we have shown that the superconducting 
transition temperature $T_c$ rises from its unstrained value of 1.5~K and peaks 
strongly at $\sim$3.5~K at a compressive strain of 
$\approx$0.6~\%~\cite{Steppke2017}.
While it is tempting to associate this with the occurrence of a Lifshitz 
transition, measurements reported to date were based only on the study of the 
diamagnetic susceptibility, and did not in themselves constitute proof that 
such a transition had occurred.
For example, the peak could also correspond to a transition into a magnetically 
ordered state induced around the Lifshitz transition~\cite{Liu2017}.
To further investigate both the existence of a Lifshitz transition and its 
consequences, we report here simultaneous magnetic susceptibility and 
electrical resistivity measurements on single crystals of Sr$_2$RuO$_4$ under 
$\langle 100 \rangle$ compressive strains of up to 1\%, and temperatures between 
1.2 and 40~K. 

\begin{figure}[b]
  \centering
  \includegraphics{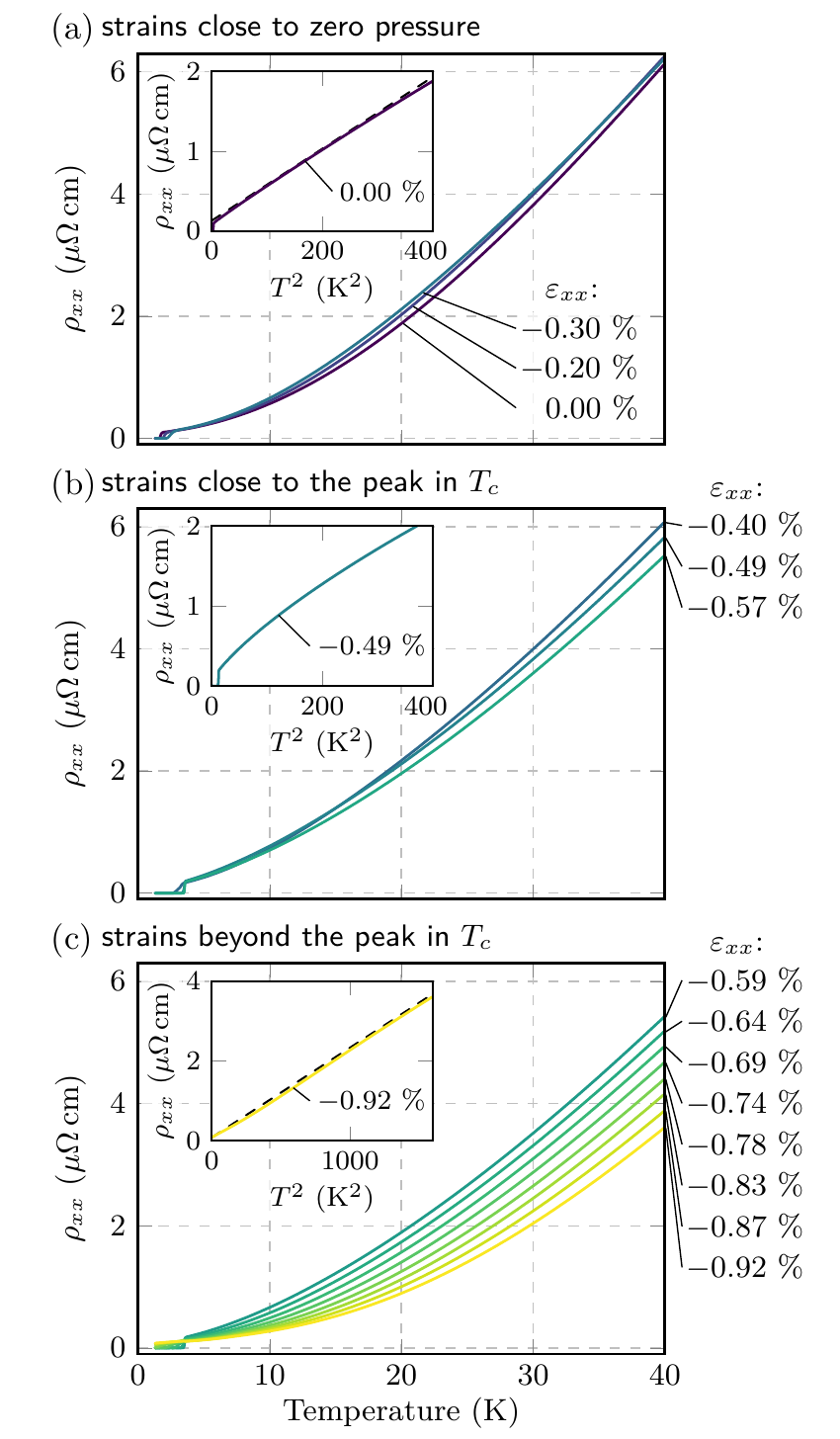}
  \caption{\label{Fig3}(color online). Temperature dependence of the resistivity 
$\rho_{xx}$ at a variety of [100] compressive strains: (a) low strains where 
quadratic temperature dependence is still observed below $\sim$20~K, (b) strains 
close to the peak in $T_c$ where the strongest deviations from $T^2$ resistivity 
are observed, and (c) the highest strains measured showing the larger extent of 
the $T^2$ region of resistivity and the rapid strain dependence of its weight.
  }
\end{figure}

A schematic of our experimental apparatus and a photograph of a crystal mounted 
for resistivity measurements are shown in Fig.~\ref{Fig2}(b).
The resistivity $\rho_{xx}$ is measured in the same direction as the applied 
pressure.
Simultaneous measurements of magnetic susceptibility were performed using a 
detachable drive and pickup coil on a probe that could be moved into place 
directly above the sample without disturbing the contacts for the resistivity 
measurements.
We rely exclusively on the susceptibility measurements to determine $T_c$, to 
avoid being deceived by percolating, higher-$T_c$ current paths.
Resistivity and susceptibility were measured using standard a.c.\ methods at 
drive frequencies between 50\,--\,500~Hz and 0.1\,--\,10~kHz respectively, in a 
$^4$He cryostat in which cooling was achieved via coupling to a helium pot that 
could be pumped to reach its base temperature of 1.2~K. 
Uniaxial pressure was applied by appropriate high voltage actuation of the 
piezo stacks shown in Fig.~\ref{Fig2}(b), using procedures described in 
Refs.~\cite{Hicks2014,Hicks2014_2,Steppke2017,Brodsky2017,Barber2017}.
After some slipping of the sample mounting epoxy during initial compression, 
all resistivity data repeated though multiple strain cycles, indicating that 
the sample remained within its elastic limit.
Two samples were studied to ensure reproducibility; further details are given 
in~\cite{Supplementary}.

In Fig.~\ref{Fig3} we show $\rho_{xx}(T)$ at various applied compressions.
Consistent with the high $T_c$ of 1.5~K at zero strain, the residual 
resistivity $\rho_{\text{res}}$ is less than 100~n$\Omega$\,cm, corresponding to 
an impurity mean free path $\ell$ in excess of 1~$\mu$m~\cite{Mackenzie1998}.
The well-established $\rho_{\text{res}} + AT^2$ 
dependence~\cite{Maeno1997,Hussey1998} is seen below $\sim$20~K in the 
unstrained sample (Fig.~\ref{Fig3}(a) and inset). 
As the strain $\varepsilon_{xx}$ increases to 0.2~\% the quadratic temperature 
dependence is retained but $A$ increases, qualitatively consistent with the 
increase in density of states expected on the approach to a Van Hove 
singularity.
Further increase of the strain results in the resistivity reaching a maximum 
and deviating significantly from a quadratic temperature dependence, as shown 
both in the main plot of Fig.~\ref{Fig3}(b) and in the inset.
As the strain is increased further, the drop in $T_c$ is accompanied by a fall 
of the resistivity, simultaneous with a recovery of the $\rho_{\text{res}} + 
AT^2$ form and a drop of the $A$ coefficient.
By $\varepsilon_{xx}$ = $-$0.92~\% $T_c$ has fallen to below 1.2~K, the 
resistivity remains almost perfectly quadratic to over 30~K, and $A$ has 
dropped to $\sim$40~\% of its value in the unstrained material 
(Fig.~\ref{Fig3}(c) and inset).

As noted above, one mechanism by which the peak in $T_c$ might not correspond 
to the Van Hove singularity is if it is cut off by a different order promoted 
by proximity to the Van Hove singularity.
This is the prediction of the functional renormalization group calculations on 
uniaxially pressurized Sr$_2$RuO$_4$ of Ref.~\cite{Liu2017}, which predict 
formation of spin density wave order.
However, the data shown in Fig.~\ref{Fig3} give no evidence for any instability 
other than superconductivity across the range of pressures studied.
There is no indication of any transition in any of the $\rho_{xx}(T)$ curves, 
before or after the peak in $T_c$.
Also, $\rho_{xx}$ falls on the other side of the peak, whereas especially at 
low temperature the opening of a magnetic gap should generally cause resistivity 
to increase.

\begin{figure}[b]
  \centering
  \includegraphics{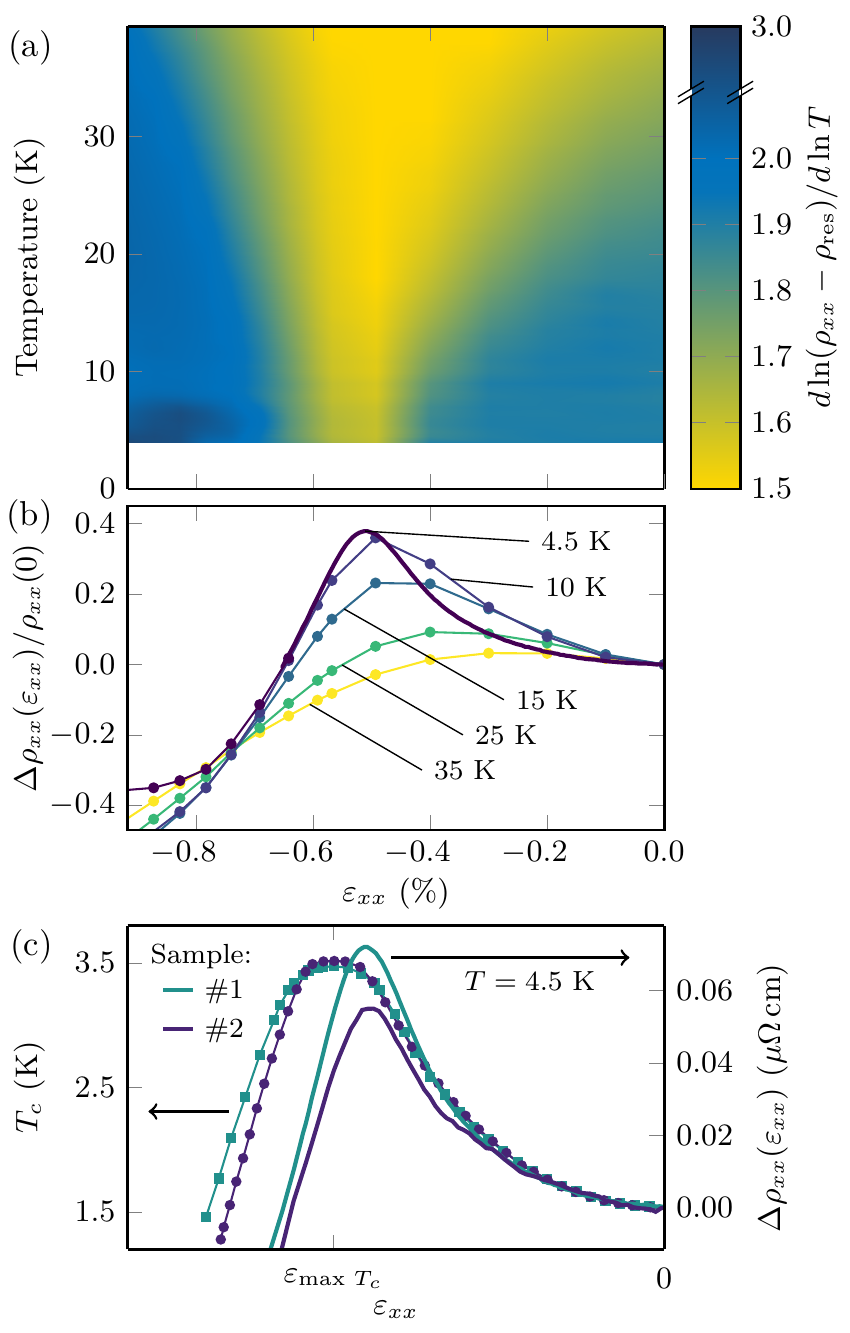}
  \caption{\label{Fig4}(color online). (a) Resistivity temperature exponent 
$\delta$ plotted against temperature and strain.
  $\rho_{\text{res}}$ was first extracted from fits of the type $\rho = 
\rho_{\text{res}} + B T^{\delta}$ and then $\delta$ was calculated as a 
function of temperature by $d\ln(\rho-\rho_{\text{res}})/d\ln 
T$~\cite{Supplementary}.
  (b) Elastoresistance at various temperatures.
  Values were calculated by interpolating between separate temperature ramps at 
a series of constant strains (Fig.~\ref{Fig3}), except for 4.5~K where the 
strain was swept continuously up to $\varepsilon_{xx} \approx$ 0.7~\%.
  (c) Comparison of the strain dependence of $T_c$ measured by magnetic 
susceptibility and the resistivity enhancement under continuous strain tuning 
at 4.5~K.
  Two samples are shown with $\rho_{\text{res}}$ of 80 and 20~n$\Omega$\,cm in 
which $\rho_{xx}$ rises to 190 and 95~n$\Omega$\,cm respectively at 4.5~K.
  In panel (c) the strain scales have been normalised to the peak in $T_c$.
  $\varepsilon_{xx}$ at the peak in $T_c$ is $-$0.56~\% and $-$0.59~\% for 
samples 1 and 2, respectively, and this difference is within our error for 
determining sample strain.
  }
\end{figure}

To correlate features of the resistivity with the evolution of the 
superconductivity more precisely, we plot, in Fig.~\ref{Fig4}, two key 
quantities associated with the resistivity and show how they compare with the 
strain dependence of $T_c$.
In Fig.~\ref{Fig4}(a) we show a logarithmic derivative plot that gives an 
indication of the strain-dependent power $\delta$ associated with a postulated 
$\rho_{\text{res}} + BT^{\delta}$ temperature dependence.
Constructing such a plot involves assumptions~\cite{Supplementary} but it gives 
a first indication of the behaviour of the resistivity and shows that $\delta$ 
drops from 2 at low and high strains to $\sim$1.5 at $\varepsilon_{xx}$ = 
$-$0.5~\%.
In Fig.~\ref{Fig4}(b) we plot the results of a measurement of the resistivity 
measured under continuous strain tuning at 4.5~K (chosen to be 1~K higher than 
the maximum $T_c$, to be free of any influence of superconductivity).
$\rho_{xx}$ is also maximum at $\varepsilon_{xx} \approx -0.5$\%.
In Fig.~\ref{Fig4}(c) we plot $T_c$ and $\rho_{xx}(T=4.5 \mathrm{~K})$ against 
$\varepsilon_{xx}$ for this sample and for a second sample with a slightly 
lower residual resistivity.
The magnitude of the resistivity increase is approximately the same for both 
samples, and for both the resistivity peaks at a slightly lower compression 
than $T_c$.

Taken together, we believe that the data shown in Figs.~\ref{Fig3} and 
\ref{Fig4} give strong evidence that we have successfully traversed the VHS in 
Sr$_2$RuO$_4$ by applying uniaxial pressure.
This is not the first time that this VHS has been reached by some form of 
tuning; in fact it has previously been traversed in the (Sr,Ba,La)$_2$RuO$_4$ 
system by explicit chemical doping of La$^{3+}$ onto the Sr 
site~\cite{Kikugawa2004,Shen2007} and by using epitaxial thin film techniques 
to grow biaxially strained stoichiometric Sr$_2$RuO$_4$ and Ba$_2$RuO$_4$ thin 
films~\cite{Burganov2016}.
In both cases, a drop of resistive exponent $\delta$ to approximately 1.4 was 
reported~\cite{Kikugawa2004,Burganov2016}.
The novelty of our results is that they are observed in crystals with such low 
levels of disorder.
In the previous experiments on the ruthenates, the inelastic component of the 
resistivity has been approximately the same magnitude by 30~K as the residual 
resistivity~\cite{Kikugawa2004,Burganov2016}, while here it is a factor of 
forty larger.
Combined with the facts that the data shown in Figs.~\ref{Fig3} and \ref{Fig4} 
cover a full decade of temperature above the maximum $T_c$, and that the Fermi 
surface of Sr$_2$RuO$_4$ is well known, we believe that this means that our 
results can set an experimental benchmark for testing theoretical understanding 
of the evolution of transport properties around an externally tuned Lifshitz 
transition.
A full treatment of the problem will require further theoretical work that is 
beyond the scope of this paper, but we close with a discussion of what is known 
so far and the extent to which it applies to our results.

\begin{figure}[t]
  \centering
  \includegraphics{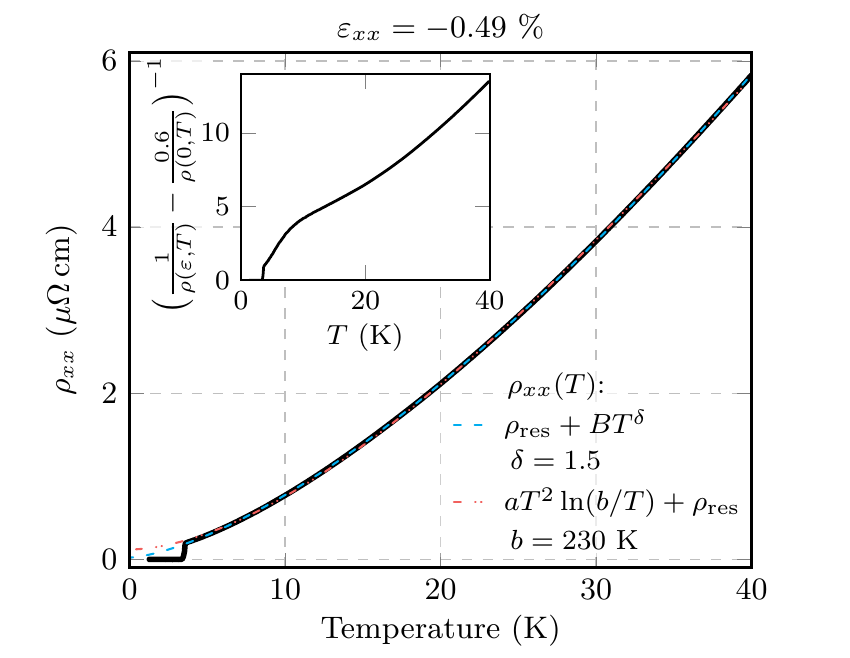}
  \caption{\label{Fig5}(color online). A comparison of different fitting 
functions for the temperature dependence of the resistivity at 
$\varepsilon_{xx}=-0.49$~\%.
  Both fits are made between 4~K and 40~K.
  The inset shows the same resistivity curve after subtracting 60~\% of the 
zero strain conductivity, estimated to be the contribution of the $\gamma$ band 
if the scattering rate of the $\alpha$ and $\beta$ sheet carriers is unaffected 
by the traversal of the Lifshitz point on the $\gamma$ sheet.
  }
\end{figure}

The effect on resistivity of traversing a Van Hove singularity has been studied 
in idealized single band models, taking into account the energy dependence of 
the density of states, electron-electron Umklapp processes and impurity 
scattering.
Depending on the form postulated for the density of states, variational 
calculations using Boltzmann transport theory in the relaxation time 
approximation have discussed resistivities of the form  $\rho(T) =  
\rho_{\text{res}} + bT^2ln(c/T)$ or $\rho(T) = \rho_{\text{res}} + 
bT^{3/2}$~\cite{Hlubina1996}.
Within experimental uncertainties, these two possibilities cannot be 
distinguished, see Fig.~\ref{Fig5}.
Numerical calculations going beyond the relaxation time 
approximation~\cite{Buhmann2013,Buhmann2013_2} predict $\delta < 2$.
The amount by which $\delta$ is reduced depends on the degree of nesting of the 
Fermi surface; $\delta = 1$ is predicted for perfect nesting.

It is interesting to note that tuning to a VHS is evidently not sufficient to 
obtain the $T$-linear resistivity that is often associated with quantum 
criticality.
Although the resistivity is enhanced in the vicinity of the VHS, $\rho(T)$ does 
not increase at nearly the rate seen for systems with $T$-linear 
resistivity~\cite{Bruin2013}.

Although the calculated temperature dependences fit the data well, we caution 
that it is questionable whether they are even applicable to Sr$_2$RuO$_4$.
As illustrated in Fig.~\ref{Fig2}(a), it is not a single-band, but a three-band 
metal, and both the tight-binding calculation presented here and full 
first-principles calculations~\cite{Steppke2017} show that the Lifshitz 
transition occurs on the $\gamma$ Fermi surface sheet.
In contrast, the $\alpha$ and $\beta$ sheets show almost no distortion at these 
strains. 
At zero strain the average Fermi velocities of each sheet are known, so for a 
sheet-independent scattering rate it is straightforward to estimate that the 
$\alpha$ and $\beta$ sheets contribute over 60~\% of the conductivity.
Under the postulate that the scattering rate of the $\alpha$ and $\beta$ sheet 
carriers is unaffected by the traversal of the Lifshitz point on the $\gamma$ 
sheet, the contribution of the $\gamma$ sheet to the resistivity at $-$0.49~\% 
strain is shown in the inset to Fig.~\ref{Fig5}, and seen to be qualitatively 
very 
different from any single-band prediction.
The likely implication of this analysis is that the scattering rate changes 
induced by the change in shape of the $\gamma$ sheet affect both the $\alpha$ 
and $\beta$ sheet carriers just as strongly as those on the $\gamma$ sheet.
However, it seems far from obvious that this should automatically be the case, 
and it would be very interesting to see full multi-band calculations for 
Sr$_2$RuO$_4$ to assess the extent to which it can be understood using 
conventional theories of metallic transport.

In conclusion, we believe that the results that we have presented in this paper 
represent an experimental benchmark for the effects on resistivity of 
undergoing a Lifshitz transition against a background of very weak disorder.
Our results stimulate further theoretical work on this topic, and highlight the 
suitability of uniaxial stress for probing this class of physics.

\begin{acknowledgments}
We thank J.\ Schmalian, E.\ Berg, and M.\ Sigrist for useful discussions and H. 
Takatsu for sample growth.
We acknowledge the support of the Max Planck Society and the UK Engineering and 
Physical Sciences Research Council under grants EP/I031014/1 and EP/G03673X/1. 
Y.M.\ acknowledges the support by the Japan Society for the Promotion of 
Science Grants-in-Aid for Scientific Research (KAKENHI) JP15H05852 and 
JP15K21717.
\end{acknowledgments}

%

% Supplemental materials
\onecolumngrid
\clearpage
\begin{center}
\textbf{\large Supplemental Materials for:\\Resistivity in the Vicinity of a Van 
Hove Singularity: \texorpdfstring{Sr$_2$RuO$_4$}{Sr2RuO4} Under Uniaxial 
Pressure}\\[2.5ex]

M.\,E. Barber,\textsuperscript{1,2} A.\,S. 
Gibbs,\textsuperscript{1} Y. Maeno,\textsuperscript{3} A.\,P. 
Mackenzie,\textsuperscript{1,2} and C.\,W. Hicks\textsuperscript{2}\\[1ex]

{\small \textsuperscript{1} \textit{Scottish Universities Physics Alliance, 
School of Physics and Astronomy,\\ University of St.\ Andrews, St.\ Andrews KY16 
9SS, U.K.}\\
\textsuperscript{2} \textit{Max Planck Institute for Chemical Physics of Solids, 
N{\"o}thnitzer Str.\ 40, 01187 Dresden, Germany}\\
\textsuperscript{3} \textit{Department of Physics, Graduate School of Science, 
Kyoto University, Kyoto 606-8502, Japan}
}
\end{center}
\setcounter{equation}{0}
\setcounter{figure}{0}
\setcounter{table}{0}
\setcounter{page}{1}
\makeatletter
\renewcommand{\theequation}{S\arabic{equation}}
\renewcommand{\thefigure}{S\arabic{figure}}
\renewcommand{\thetable}{S\arabic{table}}
\makeatother

\section{I.\texorpdfstring{\quad}{~} Details of the experimental technique}

The piezoelectric based uniaxial pressure device used in this work was first 
described in~\cite{Hicks2014_2,Hicks2014} and the later modifications 
in~\cite{Steppke2017,Brodsky2017,Barber2017}.
For this technique the samples are first prepared as long thin bars using a 
wire lapping saw and mechanical polishing.
The sample is secured into the device using epoxy, holding the sample only by 
its ends and spanning across an adjustable gap in the device.
When the two ends of the sample are pushed closer together or further apart to 
compress or tension the sample respectively, the central portion of the sample, 
where we measure its resistivity and susceptibility, is free to expand or 
contract as governed by the sample's Poisson's ratio.
This means the stress, or equivalently the pressure, in the region of the 
sample where we measure is uniaxial but not the strain.
However, we have better knowledge of the applied displacement and therefore the 
strain along the pressure axis.
We measure the displacement applied between the two ends of the sample using a 
parallel plate capacitor, in line with, and beneath the sample.
To calculate the strain we need to know the length of sample this displacement 
is applied to.
This is not trivial and leads to the largest uncertainty in calculating the 
strain in the sample.
Because the sample is held to the device using an epoxy which is relatively 
soft, the strained length of the sample is not the exposed length of sample 
between the two mounts, rather some of the epoxy in the mounts deforms too and 
the strained length is slightly longer.
Exactly how much longer depends on the sample and epoxy geometry, and their 
elastic constants.
We estimate the strained length using finite element simulations, the results 
of which and the relevant dimensions of the samples are presented in table S1.
We used the elastic constants of Sr$_2$RuO$_4$ from~\cite{Paglione2002} and 
estimate the shear modulus of the epoxy, Stycast 2850FT, to be 
6~GPa~\cite{Hicks2014}.
Because of the uncertainties in some of these values, all strains quoted have 
an uncertainty of $\sim$20~\% (a systematic error affecting all measured 
strains equally). 

\begin{table}[ht]
\caption{\label{TableS1}
Relevant sample and epoxy dimensions for calculating the strain transmission to 
the sample through the epoxy. 
$w$ and $t$ are the width and thickness of the sample.
$L_\mathrm{gap}$ is the gap between the sample plates, the exposed length of 
sample.
$d_\mathrm{epoxy}$ is the depth of epoxy above and below the sample joining to 
the sample plates.
$L_\mathrm{eff}$ is the calculated effective length of the sample as described 
in the text.
$\varepsilon_{xx,\text{ peak } T_c}$ is the strain at which the peak in $T_c$ 
was observed for each sample.
}
\begin{ruledtabular}
\begin{tabular}{cccccccc}
 sample number & growth & $w$ ($\mu$m) & $t$ ($\mu$m) & $L_\mathrm{gap}$ 
($\mu$m) & $d_\mathrm{epoxy}$ ($\mu$m) & $L_\mathrm{eff}$ ($\mu$m) & 
$\varepsilon_{xx,\text{ peak } T_c}$ (\%) \rule[-1.1ex]{0pt}{0pt}\\
 \hline
  1 & C362 & 320 & \hspace*{5pt}90 & 1100 & $\approx$25 & 1510 & $-$0.56 
\rule{0pt}{2.6ex}\\
  2 &   A1 & 310 & 100 & 1000 & $\approx$25 & 1430 & $-$0.59 \\
\end{tabular}
\end{ruledtabular}
\end{table}

The samples used in this study come from two different batches.
Each was prepared with six electrical contacts using the standard recipe, 
DuPont 6838 silver paste baked at 450$^{\circ}$C for 5 minutes, before they 
were mounted into the pressure device.
After mounting, two concentric coils were lowered above the sample to measure 
the diamagnetic signal of superconductivity by measuring the change in mutual 
inductance between the coils.

\section{II.\texorpdfstring{\quad}{~} Additional data}
\begin{figure}[t]
  \centering
  \includegraphics{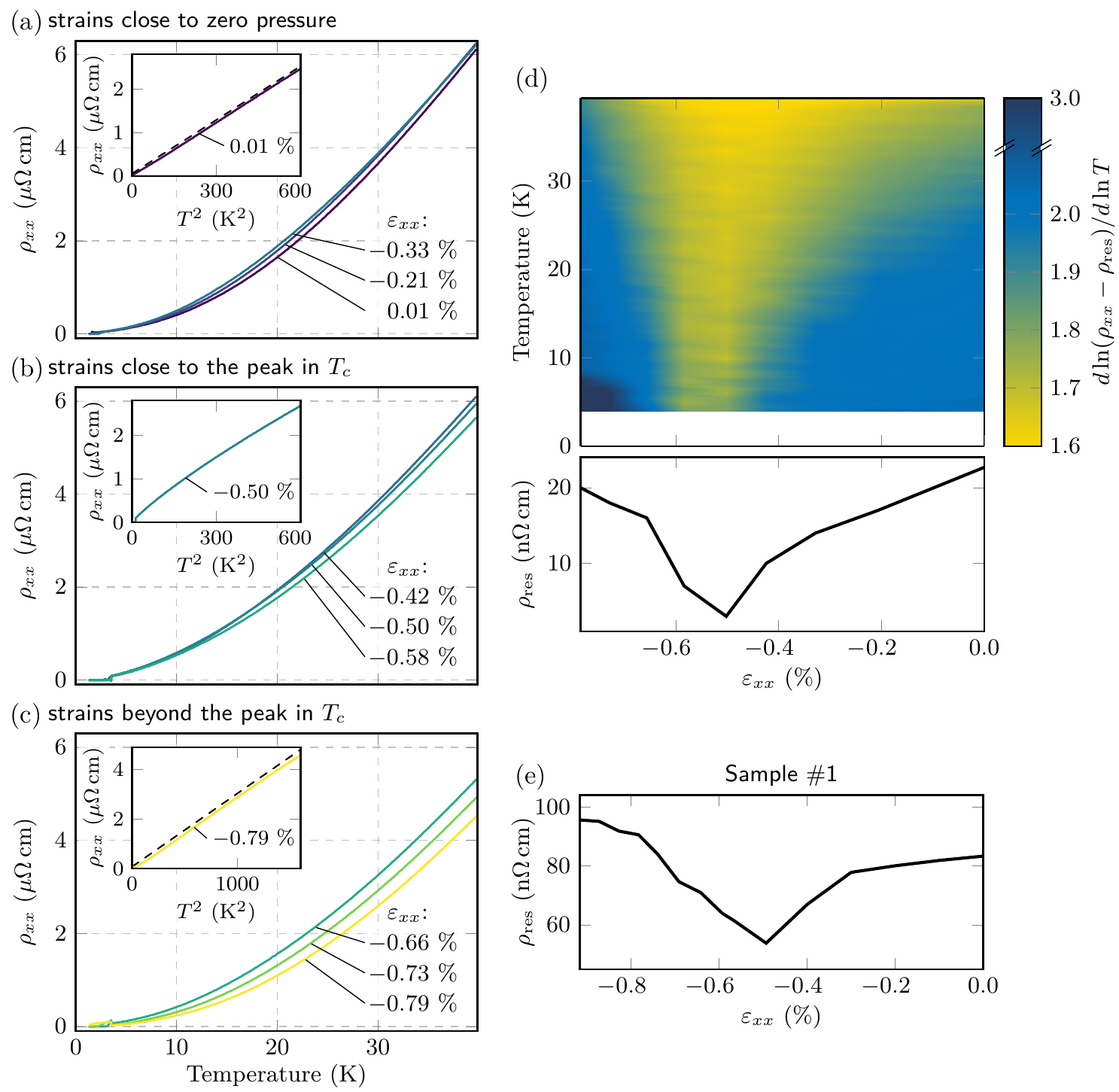}
  \caption{\label{FigS1}Additional resistivity data. Panels (a), (b), and (c) 
show the resistivity temperature dependence of the second sample in the same 
three regions across the phase diagram as described in Fig.~\ref{Fig3}. Panel 
(d) shows the logarithmic derivative plot for the second sample, depicting the 
resistivity temperature exponent $\delta$ as calculated in Fig.~\ref{Fig4}, and 
the residual resistivity used for extracting the exponent. Panel (e) shows the 
residual resistivity of sample 1 used to generate Fig.~\ref{Fig4} in the main 
text.
  }
\end{figure}

The resistivity data for sample 1 are shown in the main text.
The data for sample 2 are shown in Fig.~\ref{FigS1}.
Note that the difference in strain scale is most probably not an intrinsic 
sample-to-sample variation, and rather comes from the experimental uncertainty 
in measuring the applied strain.

Additionally in Fig.~\ref{FigS1}, the strain dependence of the fitted residual 
resistivity is shown explicitly for both samples.
To generate the temperature exponent colour maps in Figs.~\ref{Fig4}(a) and 
\ref{FigS1}(d) the residual resistivity must be subtracted from the data before 
differentiating.
To extrapolate below $T_c$ a fit of the form $\rho = \rho_{\text{res}} + B 
T^{\delta}$ was used and the range of the fit was chosen self consistently so 
that an approximately constant exponent is obtained over the temperature range 
of the fit.
Although the exact position of the crossover from $T^2$ to a lower power is 
slightly sensitive to the details of this fitting the overall picture is 
unchanged.
Similarly, although the fitted residual resistivity has a weak strain 
dependence if left as a free parameter, it is so small for both these samples 
that fixing it to be strain-independent makes no qualitative difference to the 
logarithmic derivative plots or the value of $\delta$ deduced for 
$\varepsilon_{xx} \sim -0.5$~\%.

\begin{figure}[ht]
  \centering
  \includegraphics{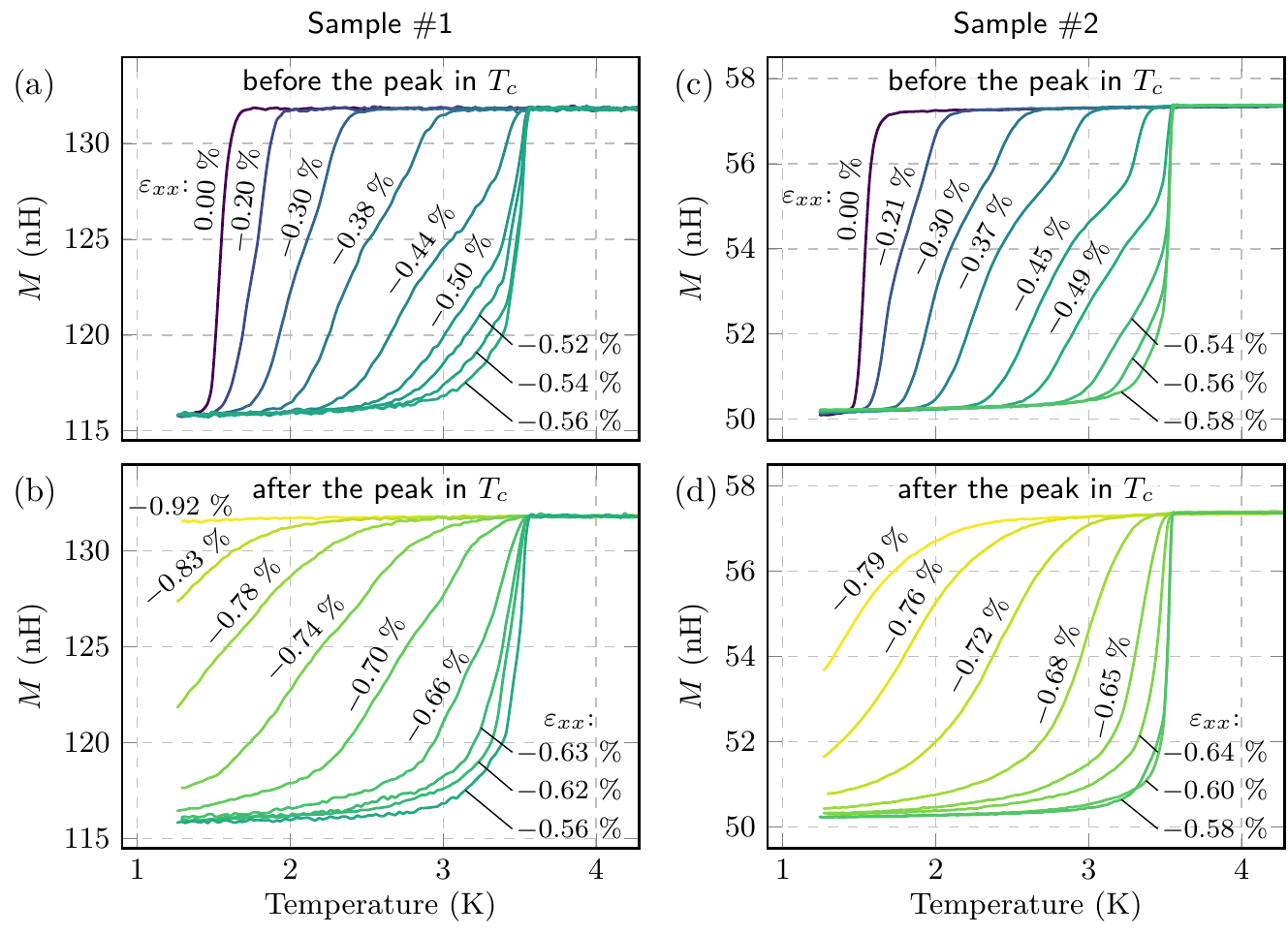}
  \caption{\label{FigS2}A.c. magnetic susceptibility measurements of both 
samples at various strains $\varepsilon_{xx}$ used for identifying 
$T_c(\varepsilon_{xx})$. (a) and (c) show measurements below the peak in $T_c$ 
for samples 1 and 2 respectively, (b) and (d) show measurements above the peak 
in $T_c$. $M$ is the mutual inductance between the two coils of the 
susceptibility sensor.
  }
\end{figure}

Fig.~\ref{Fig4} in the main text compares two curves of the strain dependence 
of $T_c$ with the resistivity at 4.5~K.
$T_c$ was obtained from temperature dependent a.c.\ magnetic susceptibility 
$\chi'(T)$ measurements and taken to be the midpoint of the transition.
Fig.~\ref{FigS2} shows the susceptibility measurements for both samples at a 
variety of strains.

\section{III.\texorpdfstring{\quad}{~} Details of the tight binding simulations}
Fig.~\ref{Fig1} in the main text presents an illustrative tight-binding 
simulation of the different effects of biaxial and uniaxial pressure.
We start from the simplest 2D nearest and next-nearest neighbour tight-binding 
model
\begin{equation}
 \Delta E(\textbf{k}) = - 2 t [cos(k_x a) + cos(k_y b)] - 2 t' [cos(k_x a + k_y 
b) + cos(k_x a - k_y b)] \,,
\end{equation}
where $a$ and $b$ are the lattice constants in the $x$ and $y$ directions.
We set the ratio $t'/t=0.3$ and work at half filling.

Under biaxial pressure, changes in the Fermi surface shape can only occur if 
the ratio of $t'/t$ changes.
Generally it is expected that $t'$ will increase faster than $t$ under 
compression~\cite{Burganov2016}.
For our simulations we make the simplest assumption that the hoppings depend 
linearly on lattice strain and make the next-nearest neighbour hopping increase 
20~\% faster with strain than the nearest neighbour hopping.
Biaxial pressure causes the unit cell lattice vector $a$ to change as
\begin{equation}
 a(\varepsilon)=b(\varepsilon)=a_0(1+\varepsilon), \qquad
 \varepsilon=\dfrac{1-\nu_{xy}}{E} \sigma,
\end{equation}
where $\sigma$ is the applied stress and the resultant strain $\varepsilon$ 
depends on the Young's modulus $E$ and Poisson's ratio $\nu_{xy}$.
We set the strain dependence of the hopping terms to
\begin{equation}
 t(\varepsilon)=t_0(1-\alpha\varepsilon) \, \qquad
 t'(\varepsilon)=t'_0(1-1.2\alpha\varepsilon).
\end{equation}
were $\alpha$ is an adjustable parameter that scales the effect of strain.

Equal uniaxial pressures cause much larger changes to the Fermi surface due to 
the much larger distortion of the unit cell as it is able to relax orthogonal 
to the pressure axis according to Poisson's ratio.
For the simulation we keep the same starting parameters as the biaxial 
simulation but now allow the hoppings to change anisotropically.
The unit cell deforms as
\begin{equation}
 a(\varepsilon)=a_0(1+\varepsilon_{xx}), \qquad
 b(\varepsilon)=a_0(1-\nu_{xy}\varepsilon_{xx}), \qquad
 \varepsilon_{xx}=\sigma_{xx}/E,
\end{equation}
and the hoppings are now different along the $x$ and $y$ directions
\begin{equation}
 t_x(\varepsilon)=t_0(1-\alpha\varepsilon_{xx}), \qquad
 t_y(\varepsilon)=t_0(1+\alpha\nu_{xy}\varepsilon_{xx}), \qquad
 t'(\varepsilon)=t'_0(1-\alpha(1-\nu_{xy})\varepsilon_{xx}/2).
\end{equation}

In the case for Sr$_2$RuO$_4$, Fig.~\ref{Fig2} in the main text, we start from 
the tight-binding parametrisation derived from fits to experimental 
data~\cite{Bergemann2003} which sets the ratios of each of the hoppings, but 
rather than using DFT calculations to set the magnitude we use the mass 
renormalisation from~\cite{Shen2007}.
We use the measured Poisson's ratio of 0.39~\cite{Paglione2002} and scale the 
hopping strengths in the same manner as for the uniaxial case above while 
keeping the total electron count constant.

\end{document}